\documentclass[aps,prb,twocolumn,showpacs,superscriptaddress,amssymb]{revtex4}
\usepackage{graphicx}
\usepackage{amsmath}
\usepackage{hyperref}
\usepackage{natbib}
\usepackage{bm}
\begin{document}

\title{Vibrational modes and lattice distortion of a nitrogen-vacancy center in diamond from 
first-principles calculations}

\author{Jianhua Zhang}
\affiliation{Ames Laboratory, Iowa State University, Ames IA 50011, USA}
\affiliation{Department of Physics and Institute of Theoretical Physics and Astrophysics, Xiamen University, Xiamen 361005, China}
\author{Cai-Zhuang Wang}
\affiliation{Ames Laboratory, Iowa State University, Ames IA 50011, USA}
\author{Z. Z. Zhu}
\affiliation{Department of Physics and Institute of Theoretical Physics and Astrophysics, Xiamen University, Xiamen 361005, China}
\author{V. V. Dobrovitski}
\affiliation{Ames Laboratory, Iowa State University, Ames IA 50011, USA}

\date{\today}
\begin{abstract}
We investigate vibrational properties and lattice distortion of negatively charged nitrogen-vacancy (NV)
center in diamond. Using the first-principles electronic structure calculations, we show that the
presence of NV center leads to appearance of a large number of quasilocalized vibrational modes (qLVMs)
with different degree of localization.
The vibration patterns and the symmetries of the qLVMs
are presented and analyzed in detail for both ground and excited orbital states of the NV center. 
We find that in the high-symmetry ($C_{3v}$) excited orbital state a pair of degenerate qLVMs becomes
unstable, and the stable excited state has lower ($C_{1h}$) symmetry. This is a direct indication of the Jahn-Teller 
effect, and our studies suggest that dynamical Jahn-Teller effect in the weak coupling regime takes place. 
We have also performed a detailed comparison of our results with the available
experimental data on the vibrations involved in optical
emission/absorption of the NV centers. We have directly demonstrated that, among other
modes, the qLVMs crucially impact the optical properties of the NV centers in 
diamond, and identified the most important groups of qLVMs. Our results are important for
deeper understanding of the optical properties and the orbital relaxation associated with 
lattice vibrations of the NV centers.
\end{abstract}


\maketitle

\section{Introduction}

The negatively charged nitrogen-vacancy (NV) impurity centers in diamond
have attracted much interest in recent years. They demonstrate a uniquely
favorable combination of spin and optical properties: a NV electronic spin
has long coherence time \cite{Childress06,Gaebel06,Hanson08,Balasubramanian09}, 
the spin state of a single NV center
can be optically initialized and read out \cite{Gruber97,Jelezko02}, and can be manipulated
both optically \cite{Santori06,BuckleyAwsch10,Robledo10} and magnetically \cite{Fuchs09,Jelezko04}. As a result, NV centers
constitute promising candidates for applications in quantum information
processing \cite{Childress06PRL,JelezkoGate04,Dutt07},
high-sensitivity magnetometry with nanoscale
resolution \cite{Taylor08,Balasubramanian08,deLangeMagnetometry10}, 
and photonics \cite{Beveratos02,Kurtsiefer00,Babinec10,Kaiser09,Togan10}. The NV centers also present an
excellent platform for studying such fundamental problems of quantum
mechanics as the dynamics of quantum spins coupled to their environment
\cite{Childress06,Hanson08,deLange10,Ryan10,Naydenov10}, and for exploring quantum control 
and dynamical decoupling of solid-state spins \cite{Fuchs10,Fuchs09,deLange10,Ryan10,Naydenov10}.

The electronic, optical, and spin properties of the NV centers are strongly
affected by the lattice vibrations. For instance, due to slightly different
atomic arrangement in the ground state and the excited state of the NV center, the
optical emission and absorption is accompanied by the lattice motion, 
i.e.\ involves the vibrational lattice excitation \cite{Pekar,HuangRhys50,Nizovtsev01}.
The impact of lattice motion on the properties of the
NV centers has been under investigation since 1970s \cite{Davies,DaviesHamer} 
till now \cite{Fuchs10,Fu09,Rogers09TimeAv,Batalov09}. The lattice vibrations are responsible for the
broadening of the light emission spectrum and appearance of 
additional spectral features \cite{Davies,DaviesHamer}, for depolarization of the emitted
photons \cite{Fu09,DaviesHamer,Kaiser09}, and for the orbital and spin relaxation of the NV centers
\cite{Fu09,Rogers09TimeAv,Batalov09,Fuchs10}. The
localized and quasi-localized vibrational modes are 
especially important for defect centers in semiconductors, and are often
used for investigation of their properties \cite{27,28,29,30}. 
The detailed
studies of the lattice vibrations, and of the (quasi)localized modes
in particular, would be very useful for deeper understanding of the optical
properties of NV centers, and would greatly improve our knowledge of the
phonon-assisted spin and orbital relaxation of this system. 
Such studies are also
useful for improvement of the optical properties of NV centers, as 
required for photonic-related applications \cite{Beveratos02,Kurtsiefer00,Babinec10,Kaiser09,BuckleyAwsch10,Togan10}.
Very recently, an important first study of several localized modes has been reported
for the NV center \cite{GaliLVM11}. But detailed studies of 
other numerous localized and quasi-localized modes, which could be important for 
better understanding of the experimental data \cite{Davies,DaviesHamer}, are still lacking.

In this paper, we present a detailed first-principle study of the localized distortions and
vibrational properties of the negatively charge NV centers in diamond in its ground
and excited orbital states. We show that the
presence of the NV impurity noticeably modifies the vibrational spectrum, 
and leads to appearance of numerous quasi-localized vibrational modes
(qLVM) with different degree of localization.
We present detailed description of several important 
qLVM, and analyze the changes in these modes upon optical excitation.
Comparing our results with earlier experiments \cite{Davies,DaviesHamer},
we show that qLVM crucially
influence the luminescence properties of the NV centers. The partial
density of states (DOS) of the vibrations (quasi)localized on the NV center
is in excellent quantitative agreement with the results of Ref.~\onlinecite{Davies}. 
This is the first, to our knowledge, first-principles explanation of these experiments.
Our results also suggest that the one-phonon
sidebands in the NV center emission and absorption spectra are controlled by qLVM. We
give a detailed description of the relevant modes, and show that their properties
(symmetry and energy) are in agreement with the experimental results
\cite{DaviesHamer}. 

Another interesting finding reported here is that some of
the localized vibrational modes are unstable for the high-symmetry ($C_{3v}$)
excited orbital state of the NV center. This is a direct indication that
the Jahn-Teller (JT) effect influences the excited state, either by lowering the
symmetry of the system (static JT effect), or by entangling vibrational and
electronic states (dynamic JT effect). Our calculations, as well as the known
experimental facts, evidence the latter. According to our results, the frequencies of the relevant modes
are much larger than the JT energy, indicating the regime of the weak vibronic coupling, in
agreement with known facts (such as e.g.\ small renormalization of the $g$-factor and 
the shape of the optical transition bands).
However, detailed quantitative theory
of the JT effect in NV centers requires a separate focused investigation, which is to
be performed in the future.

The rest of the paper is organized as follows. In Sec. II we describe 
the computational method used in the paper. In Sec. III we give a
detailed description of the electronic structure of the negatively charged NV color
center, and optimization of the atomic geometry. We discuss the local
distortion of the atomic configuration in the ground and the excited states,
and discuss the possibility of the static and dynamic JT effects in the
excited state.
In Sec. IV we present the results for the lattice vibrations in the
ground and in the excited state. We discuss the phonon DOS, 
localization of different modes, and their properties. In Sec. V, we perform
and extensive comparison of
our results with the known experimental data.
Conclusions are given in Sec VI.

\section{Computational method}

The first-principles calculations were performed within the
density functional theory (DFT) under the generalized
gradient approximation (GGA) of Perdew, Burke, and Ernzerhof
(PBE) \cite{24}, including spin polarization. The Vienna Ab-initio Simulation Package (VASP) 
\cite{25} has been used to
perform geometry optimization and force calculations. A
single electron was added to the electron occupancy for the
negatively charged vacancy. The extra electron is balanced
by a uniform positive background to ensure the overall
supercell is electrically neutral. We used $\Gamma$ point for 
$k$-point sampling and a plane wave basis set with a cutoff of
420~eV.

The atomic structures of the NV center in its electronic
ground and excited states have been optimized before the
calculations for the vibration modes are performed. In the
geometry optimization calculations, a 215-atom ($3\times 3\times 3$) cubic
supercell with the box length of $3a_0 = 10.719$~\AA\ was used,
where $a_0$ is the lattice constant of the diamond structure optimized
by the VASP calculation. All internal atomic positions were
relaxed until the forces were smaller than $10^{-3}$~eV/\AA.

The vibrational modes for the ground state and excited state
of the NV center were calculated within the harmonic
approximation by first-principles calculations. The force-constant matrix of the supercell containing a negatively
charged NV center defect are evaluated through the forces
calculated via the Hellmann-Feynman theorem when each atom
in the supercell is displaced one-by-one from its
equilibrium position along $\pm x$, $\pm y$, and $\pm z$ directions
respectively with a small displacement \cite{26}. We chose a
displacement amplitude of 0.02~\AA\ from the atomic equilibrium
positions in three Cartesian directions in the force
calculation. The amplitude (0.02~\AA) is chosen here because
it is small enough to remain in the harmonic part of the
potential but large enough to avoid numerical noise. The
forces acting on all other atoms within the supercell due to
the one-by-one displacements of the atoms are used to
construct the force constant matrix of the system. The 642
phonon frequencies plus three translational modes and their
corresponding eigenvectors are obtained by diagonalizing the
$645\times 645$ force constant matrix.

\section{Electronic structure of the NV center and local distortions
from first-principle geometry optimization}

\begin{figure}
\includegraphics[width=9cm]{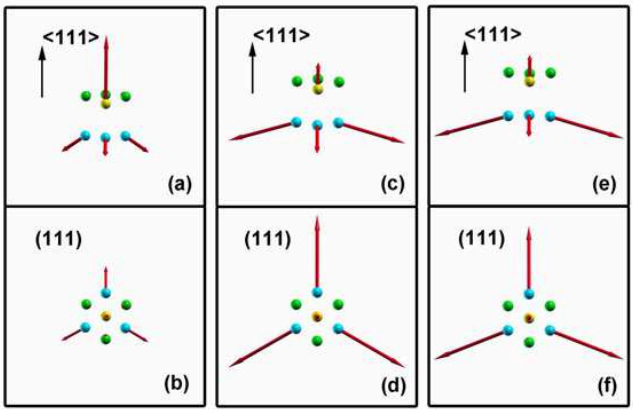}
\caption{\label{fig1} 
(Color online). Atomic displacements after geometry optimization, side view
(top panels) and top view (bottom panels) with respect to the $\langle 111\rangle$ axis. The
red arrows show the atomic displacements, the arrow length is proportional 
to the diplacement magnitude.
(a),(b): ground state, $C_{3v}$ symmetry; (c),(d): excited $^3E$ state with 
$C_{3v}$ symmetry; (e),(f): the same excited state, but fully relaxed, 
with $C_{1h}$ symmetry. Yellow spheres denote N atom, blue spheres denote the
C atoms adjacent to the vacancy, green spheres denote the C atoms adjacent to the N atom.}
\end{figure}

The nitrogen-vacancy (NV) color center consists of a substitutional
nitrogen atom (N) and an adjacent vacancy (V) in the diamond lattice. 
The main features of its atomic and electronic structure are known rather well,
and have been recently reviewed in detail in Ref.~\onlinecite{Manson06}.
For a negatively charged NV center, the six excess electrons are
localized around the center. In the orbital ground state, the atomic
arrangement has $C_{3v}$ point symmetry, where one of the $\langle 111\rangle$
crystallographic
axes (passing through the N atom and the vacancy site) 
is the symmetry axis. Our calculations of the relaxed
atomic structure are
in agreement with these facts, see Fig.~\ref{fig1}. The N and C 
atoms at the nearest-neighbor shell of the vacancy
relax outwards from vacancy. The atomic
displacements from initial unrelaxed structure for the N and
C atoms are 0.148~\AA\ and 0.101~\AA, respectively, and the distances
from the vacancy to nitrogen and carbon atoms in the relaxed structure
are 1.695~\AA\ and 1.646~\AA, respectively. These
results agree with those obtained previously, see e.g.\ Refs.~\onlinecite{Lusz04,Larsson08,Gali08,Hossain08}.

\begin{figure}
\includegraphics[width=9cm]{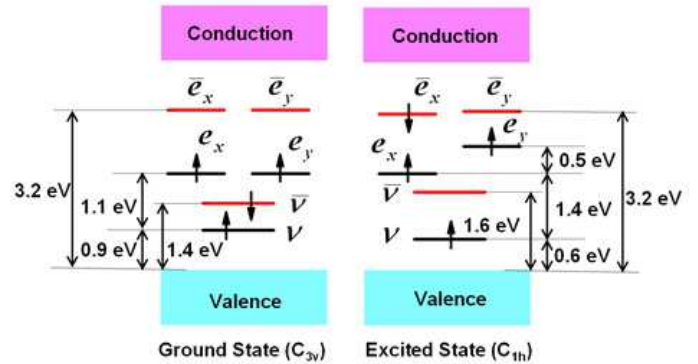}
\caption{\label{fig2} 
(Color online). The single-electron orbitals in the ground $^3A_2$ and 
excited $^3E$ orbital states of the negatively charged NV center. The energies are given with
respect to the top of the valence band. The states within the valence band are not shown.
The symbols with the bar denote the
spin-down states, the symbols without the bar correspond to the spin-up states.
The states $\nu$ and $\bar\nu$ have $a_1$ symmetry, all other states have $e$ symmetry.
We use the notations of the parent $C_{3v}$ group to classify the states of the
excited state with $C_{1h}$ symmetry, in spite of the reduced symmetry of the latter.}
\end{figure}

The $C_{3v}$ point group
has two types of non-degenerate singlet representations ($A_1$ and $A_2$),
and a doubly degenerate representation $E$; below we will use lower-case letters
to denote representations for single-electron states and vibration modes,
and upper-case for many-electron states.
The many-electron ground state of a NV center is an
orbital singlet with the total electron spin $S=1$, corresponding to
$^3A_2$ representation (spin triplet, orbital singlet of $A_2$ type). This
state can be visualized within the qualitative model \cite{Manson06} where six electrons
localized at the NV center populate the single-electron spin-resolved
orbitals \cite{LenefRand96,Manson06}, as shown in Fig.~\ref{fig2}. 
Two electrons occupy the spin-up and spin-down levels
immersed in the valence band, we exclude them from consideration. The next two
electrons occupy the states $\nu$ and $\bar\nu$ (spin-up and spin-down, respectively)
of $a_1$ symmetry. These orbitals are in the band gap of the bulk
diamond, and are localized at the carbons surrounding the vacancy and near the nitrogen.
The two remaining electrons occupy two degenerate spin-up orbitals of $e$ symmetry
($e_x$ and $e_y$); the corresponding spin-down orbitals are unoccupied. 
The energies of the single-electron states calculated here are in agreement with 
the previously published results \cite{Lusz04,Larsson08,Gali08,MaGali10}.

Upon optical excitation the NV center undergoes transition to an excited state
of $^3E$ symmetry (spin triplet, orbital doublet), which can be visualized as a
transition of an electron from $\bar\nu$ to one of the spin-down orbitals $\bar e_x$
or $\bar e_y$. Starting
from the optimized $C_{3v}$ geometry of the ground state, we found
that the system conserves this symmetry after geometry
optimization in the excited state. The outward atomic
displacements of the N and C atoms are 0.090~\AA\ and 0.156~\AA,
respectively, see Fig.~\ref{fig1} (middle panel), and the resulting
C-to-vacancy and N-to-vacancy distances are 1.68~\AA\ and
1.63~\AA, respectively, also consistent with previous work \cite{Gali08}.

However, if the geometry optimization for the excited state starts from
a low-symmetry atomic arrangement, then the fully relaxed
structure have the symmetry of the point group $C_{1h}$. The signs of
an instability of the $C_{3v}$ configuration are seen in the phonon
spectrum: the high-symmetry $C_{3v}$ excited state possesses two degenerate soft
unstable phonon modes with formally negative frequencies.
The energy of
the $C_{1h}$ atomic configuration is 6.4~meV lower than the
energy of the high-symmetry $C_{3v}$ configuration.
While the
displacements of the N atom in the $C_{1h}$ and $C_{3v}$ configurations are the same,
the displacements of the carbons adjacent to the vacancy are different:
one moves by 0.147~\AA\ away from the vacancy site, and the two others
move by 0.158~\AA, see Fig.~\ref{fig1}. To check the
possible finite-size effect in the geometry relaxation, 
we repeated this calculation for a twice larger supercell, with
511 atoms. The same result has been obtained, with the fully relaxed structure
in the excited orbital state having $C_{1h}$ symmetry, and the energy
difference with the $C_{3v}$ configuration was 7.7~meV. The slightly larger energy
difference for bigger supercell is expected, since more
atoms around the defect can be relaxed. Note that the unstable modes
have not been found in the recent calculations \cite{GaliLVM11},
either due to different calculation method, or due to smaller
size of the supercell used there.

This symmetry breaking is not very surprising. The structure of the excited
state corresponds to the typical case where the Jahn-Teller (JT) effect
arises \cite{Bersuker,DaviesReview}: the system of $C_{3v}$ symmetry with two degenerate $E$-type 
electronic states and two degenerate $e$-type phonon modes ($E\otimes e$
case). In agreement with expectations, the energy difference between the
parent high-symmetry $C_{3v}$ configuration and the symmetry-broken
$C_{1h}$ structure comes mainly from the asymmetric relaxation of
the atoms adjacent to the vacancy. Therefore, although the magnitude of the energy difference 
does not noticeably exceed the error of the DFT-PBE calculations, 
we expect that the $C_{1h}$ configuration is indeed the lowest-energy 
configuration at zero temperature {\it within adiabatic approximation\/}.
This corresponds to the static JT distortion of the NV center's
atomic structure, when the energy of the system has three potential
energy minima, related to each other via rotation by the angle $2\pi/3$
around the symmetry axis of the parent $C_{3v}$ structure.
At higher temperatures, exceeding the $C_{3v}$--$C_{1h}$ energy difference,
(above $\sim 100$~K), all energy minima would be occupied equally, and 
an effective $C_{3v}$ symmetry would be observed in most (but not all) experimental
situations \cite{Bersuker}. 

There is another possibility for the original $C_{3v}$ symmetry to be restored, the
dynamical JT effect. Our calculations have been performed within the 
adiabatic approximation, where the
atoms are treated as classical point objects with well-defined coordinates.
This approximation may become invalid when two electron states are degenerate,
as is the case for the $C_{3v}$ excited state \cite{Bersuker}.
The atomic motion then should be treated quantum-mechanically, and 
the vibrations may become strongly entangled with orbital
degrees of freedom. When the JT energy barrier $E_{JT}$ is much lower than
the relevant vibrational frequencies $\omega$, 
the relief of the potential energy surface is negligible in comparison
with the zero-point atomic vibrations, and 
the total symmetry of the coupled electron-lattice system is $C_{3v}$
even at zero temperature. Recent experiments evidence in favor of
the dynamical JT effect in NV centers \cite{Fu09}, while the static JT 
seems to contradict the existing data \cite{DaviesHamer,DaviesReview}.

Our calculations seem to support the picture of the dynamical JT effect, 
giving the barrier height $E_{JT}$ of order of
10~meV, and the relevant vibrational frequencies being an order of magnitude larger.
This is the indication of the weak coupling regime, and is in agreement with
the known experimental facts (such as e.g.\ small renormalization of the $g$-factor,
and the shape of the optical transition bands).
However, detailed quantitative theory requires a special focused investigations,
which would include the lattice nonlinearity and the many-body effects associated
with it. Such a study is to be performed in the future. 
Therefore, below we choose
the $C_{1h}$ configuration to study the lattice vibrations in the excited state.
Fortunately, the results below are not very sensitive to this choice.

\section{Lattice vibrations and quasi-localized vibrational modes of the NV center}

When a defect is introduced in a perfect lattice, the localized vibration
modes may appear in the phonon spectrum \cite{Maradudin}. If the energy of such a mode
lies in the gap of the phonon spectrum, or above the highest phonon 
mode of the perfect lattice, then the vibration amplitude decays
exponentially away from the defect (complete localization). But when this mode lies inside the phonon
spectrum of the perfect lattice, it can hybridize with the delocalized phonons,
and give rise to a quasi-localized vibrational mode (qLVM), which is not completely
localized, but whose amplitude still concentrates in the immediate vicinity of the defect.
The degree of the mode localization can be quantified by its inverse participation ratio (IPR),
defined as 
\begin{equation}
P=\sum_{j=1}^{3N} ({\bf u}_j\cdot{\bf u}_j)^2 \left[\sum_{j=1}^{3N} {\bf u}_j\cdot{\bf u}_j\right]^{-2}
\end{equation}
where $N$ is the number of atoms in the system, and ${\bf u}_j$ is the normalized 3$N$-dimensional
vector of the atomic displacements of the corresponding vibrational eigenmode \cite{WangHo}.
IPR characterizes a typical number of atoms involved in a given vibration, so that
for completely delocalized phonons $P\sim 1/N$ is almost zero for macroscopic systems, while for 
completely localized modes $P$ remains finite independently of the system size.

\begin{table*}[tbp]
\caption{\label{tab1}
Parameters of the quasi-localized vibrational modes with largest IPR in the $C_{3v}$ ground state:
frequency $\omega$, IPR $P$, and symmetry.}
\begin{ruledtabular}
\begin{tabular}{ccccccccc}
$\omega$ (meV) & 41.66 & 71.21 & 72.16 & 74.79 & 97.95 & 138.63  & 156.64 & 156.70\\
$P$ & 0.023 & 0.018 & 0.046 & 0.047 & 0.017 & 0.017 & 0.029 & 0.031 \\
Symmetry & $e$ & $e$ & $a$ & $a$ & $a$ & $e$ & $e$ & $a$ 
\end{tabular}
\end{ruledtabular}
\end{table*}

\begin{table*}[tbp]
\caption{\label{tab2}
Parameters of the quasi-localized vibrational modes with largest IPR in the $C_{1h}$ excited state:
frequency $\omega$ and IPR $P$. }
\begin{ruledtabular}
\begin{tabular}{ccccccccccccc}
$\omega$ (meV) & 31.62 & 71.53 & 73.42 & 74.82 & 97.70 & 134.07 & 156.85 & 157.34 & 162.69 & 162.77 & 163.35 & 164.57\\
$P$ & 0.017 & 0.060 & 0.028 & 0.034 & 0.016 & 0.017 & 0.022 & 0.028 & 0.037 & 0.017 & 0.037 & 0.082
\end{tabular}
\end{ruledtabular}
\end{table*}

\begin{figure}
\includegraphics[width=9cm]{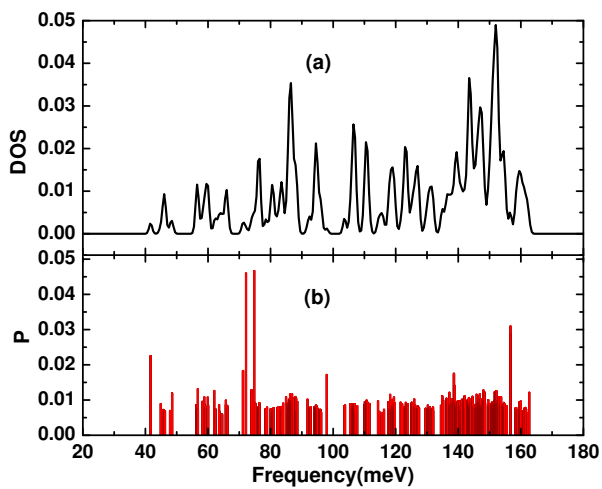}
\caption{\label{fig3} 
(Color online). The total density of vibratioinal states (a) and the IPRs of
the corresponding modes (b) for the ground state of the NV center. The graph (black solid line) 
in panel (a) was smeared, following
the standard practice, by convolution of the calculated discrete set of frequencies with 
the Gaussian line of the width~0.05 meV.}
\end{figure}

\begin{figure}
\includegraphics[width=9cm]{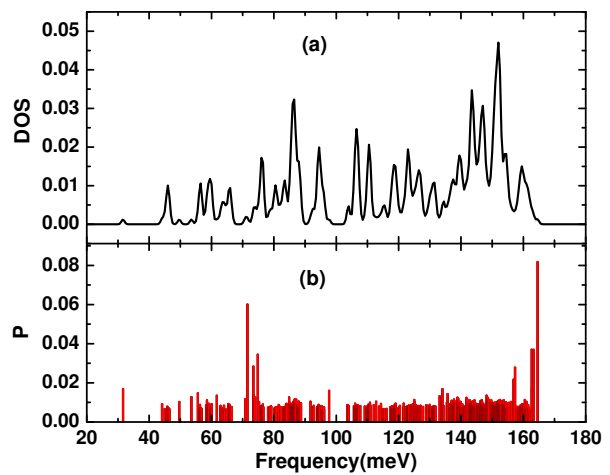}
\caption{\label{fig4} 
(Color online). The total density of vibratioinal states (a) and the IPRs of
the corresponding modes (b) for the excited $C_{1h}$ state of the NV center. The curve (black solid line) 
in panel (a) was smeared, following
the standard practice, by convolution of the calculated discrete set of frequencies with 
the Gaussian line of the width~0.05 meV.}
\end{figure}

The total density of states (DOS) of the lattice vibrations in the ground state of the NV center
and the IPRs of the corresponding modes are shown in Fig.~\ref{fig3}. 
The corresponding graphs for the
excited orbital state of the $C_{1h}$ symmetry are presented in Fig.~\ref{fig4}. 
In both Figures,
one can see a large number of modes with different degree of localization, including 
well-pronounced qLVMs with $P$ of order of 0.03--0.05. Such modes are very strongly localized
near the NV center, and involve noticeable vibrations of only 4--6 atoms; the remaining 
contribution to the value of $P$ is due to very small vibrations of more distant atoms.
The list of modes with the largest IPR is given
in Table~\ref{tab1} for the ground state and in Table~\ref{tab2} for the excited $C_{1h}$ state.
Note that all modes are infrared-active.
These results show that a large number of qLVMs, with different degree of localization,
appears in the diamond lattice with the NV center. This is in contrast with the finding
reported recently \cite{GaliLVM11}, where only six localized modes have been reported.
This might be due a different computational method, or due to a smaller size of the supercell used in  
Ref.~\onlinecite{GaliLVM11}.

Upon the optical excitation, the qLVMs undergo significant changes, as clearly seen from
comparison of Figs.~\ref{fig3}(b) and \ref{fig4}(b), and Tables \ref{tab1} and \ref{tab2}. 
The qLVM with the frequency of 41.66~meV (336~cm$^{-1}$) which
is clearly seen in the ground state, disappears in the excited state, and a new qLVM
at 31.62~meV (255~cm$^{-1}$) appears in the excited state. Also, the group of qLVMs 
with frequencies slightly above 70~meV undergoes some changes: they become less
localized in the excited state.
Also, the group of the qLVMs at high frequencies, around 140--160~meV, undergoes
noticeable change.

\begin{figure}
\includegraphics[width=9cm]{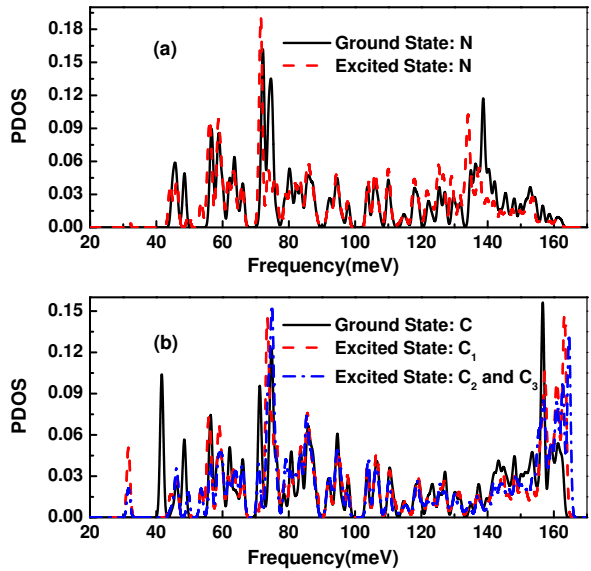}
\caption{\label{fig5} 
(Color online). (a): Projected DOS (PDOS) of the vibrations associated with the carbon atoms adjacent to the vacancy.
The ground-state PDOS of all carbons are the same (black solid line), while in the excited
state the PDOS of the C$_1$ carbon (dash-dotted green line) is different from the PDOS of the C$_2$ and C$_3$ atoms
(dashed red line). (b): PDOS of the vibrations associated with the nitrogen atom.
The ground-state PDOS is shown with black solid line, and the excited-state PDOS is shown with
the dashed red line. All curves were smeared, following
the standard practice, by convolution of the calculated discrete set of frequencies with 
the Gaussian line of the width~0.05 meV.}
\end{figure}

Note that most qLVMs are barely seen in the total DOS, Figs.~\ref{fig3}(a) and \ref{fig4}(a),
and only the IPR analysis allows us to discern them. 
To see the changes in the qLVMs in more detail, in Fig.~\ref{fig5} we present the projected DOS (PDOS)
corresponding only to the vibrations of the nitrogen and the three carbons adjacent to the vacancy.
In the ground state, all carbons are equivalent, and have the same PDOS.
In the $C_{1h}$ excited state, one carbon (denoted as C$_1$)
is different from the two others (denoted as C$_2$ and C$_3$, and related to each other by
reflection in the mirror plane \{110\} passing through N, C$_1$, and the vacancy site),
so that only the PDOS of the C$_2$ and C$_3$ atoms are the same.
Fig.~\ref{fig5} shows, in particular, that both the ground-state qLVM at 41.66~meV 
and the excited-state qLVM at 31.62~meV are associated almost completely with the carbon
vibrations. The qLVM at 41.66~meV is doubly degenerate ($e$ type), so the contributions
of different carbons are not well defined, while the excited-state qLVM at 31.62~meV is associated
exclusively with C$_1$ vibrations.

Next, the group of qLVMs slightly above 70~meV, which includes both nitrogen and carbon vibrations,
shows moderate changes in the excited state. But the changes in the high-frequency qLVMs,
around 140--160~meV,
are very noticeable: the modes associated
primarily with N vibrations shift towards lower energies, while the carbon vibrational
modes acquire higher frequencies. 
These changes are likely to come from the electron transfer from carbons to the nitrogen
(from the $\bar\nu$ to the $\bar e_x$ orbital, Fig.~\ref{fig2}),
which accompanies the optical excitation. Filling of the orbitals near carbon atoms makes the
bonds stiffer, while the depletion of the nitrogen orbitals makes them softer, with corresponding
changes in the qLVM frequencies. 

\begin{figure}
\includegraphics[width=9cm]{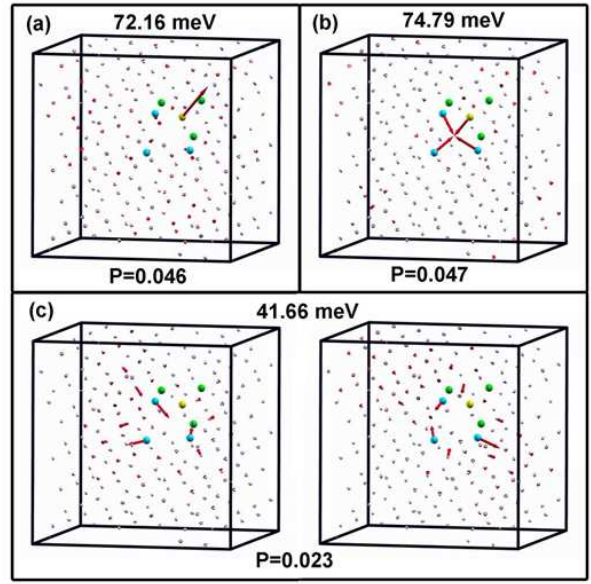}
\caption{\label{fig6} 
(Color online). qLVMs in the ground state of the NV center: atomic vibration patterns of  
several modes of different symmetry having largest IPR. The
red arrows show the atomic vibrations, the arrow length is proportional 
to the vibration amplitude of a given atom. (a) and (b) --- the modes of $a$ symmetry,
with energies of 72.16~meV and 74.79~meV,
(c) --- two degenerate modes of $e$ symmetry with the energy of 41.66~meV appearing
in the ground state but not in the excited state.
Yellow spheres denote N atom, blue spheres denote the
C atoms adjacent to the vacancy, green spheres denote the C atoms adjacent to the N atom.
Small grey speres are the other carbon atoms in the diamond lattice.}
\end{figure}

\begin{figure}
\includegraphics[width=9cm]{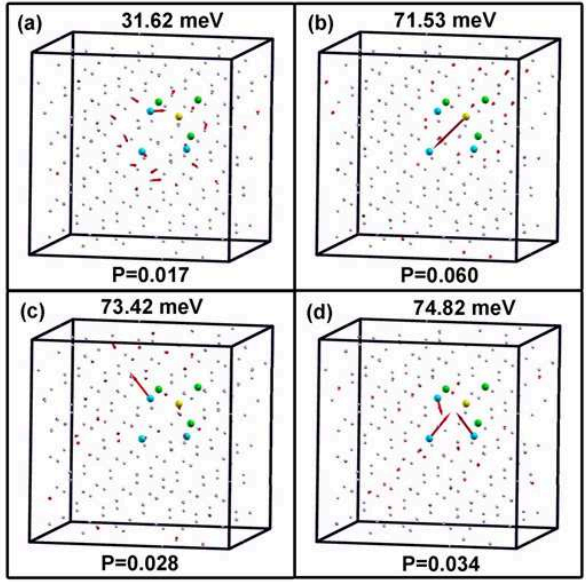}
\caption{\label{fig7} 
(Color online). qLVMs in the excited state of the NV center: atomic vibration patterns of  
several modes of different symmetry having largest IPR. The
red arrows show the atomic vibrations, the arrow length is proportional 
to the vibration amplitude of a given atom. (a) --- the mode of frequency 31.62~meV
appearing in the excited state, (b), (c), and (d) --- the modes with frequencies
71.53~meV, 73.42~meV, and 74.82~meV: these are the modes with largest IPR
belonging to the group of qLVMs near 70~meV. 
Yellow spheres denote N atom, blue spheres denote the
C atoms adjacent to the vacancy, green spheres denote the C atoms adjacent to the N atom.
Small grey spheres are the other carbon atoms in the diamond lattice.}
\end{figure}

Our calculations provide even more detailed picture of each relevant qLVM,
identifying its type (bond stretching, bond bending, etc.), and visualizing
the motion of all atoms. In Figs.~\ref{fig6} and \ref{fig7}, we present 
the atomic vibration patterns of several important qLVMs for the ground and
the excited state, respectively. In particular, the qLVMs with frequencies around 70 meV
are shown: these modes are important for discussion of experiments given below.
Detailed assessment of several other qLVMs 
is presented in the Appendix. Here we just mention that
all the analysis and the ideas presented above are consistent with detailed
consideration of qLVMs.

\section{Comparison with the experimental results.}

The optical transition between the ground and the excited state 
is accompanied by the changes in the atomic positions in the neighborhood of the defect site, 
i.e.\ involves absorption and emission of the lattice vibration quanta \cite{Pekar,HuangRhys50}.
The zero-phonon line (ZPL) in the optical spectrum, which corresponds
to the vertical transition (no absorption/emission of the vibrational quanta), is accompanied by
sidebands corresponding to absorption/emission of vibration quanta.
For NV center, the ZPL 
has the energy 1.945 eV (640 nm), while the vibronically broadened sidebands
extend well beyond 750~nm at room temperature. 
Since the optical transitions are tightly linked to the lattice vibrations, 
the studies of the optical emission and absorption provide clues about the
relevant vibrational modes. In particular, in Ref.~\onlinecite{Davies},
the absorption and luminescence spectra have been studied in detail,
and the DOS of the vibrations participating in the optical transition
has been calculated. 
It has two broad peaks with maxima around 70~meV and 140~meV
and width of order of 50~meV each, see Fig.~\ref{fig6}. Later, more detailed examination
of the PL spectra \cite{DaviesHamer} revealed the well-defined one-phonon sidebands, 
which correspond to a vibration mode of $a$ symmetry with
the energy of 70~meV. Moreover, the absorption band was different from the emission
band, demonstrating at least two closely spaced peaks.

\begin{figure}
\includegraphics[width=9cm]{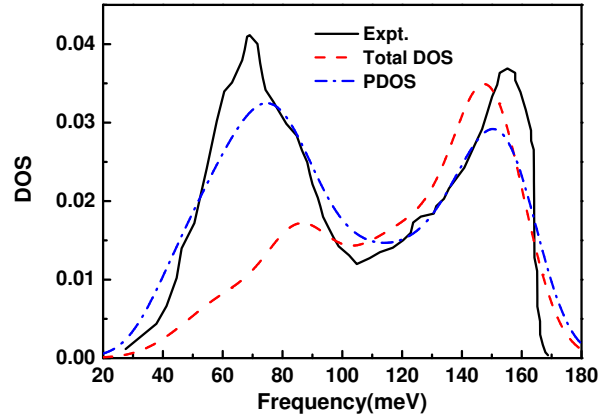}
\caption{\label{fig8} 
(Color online). The experimentally determined DOS of the vibrations affecting
the optical transition (black solid line) are compared with the calculations
results. The blue dash-dotted line denotes the
calculated projected DOS of the vibrations associated with the four atoms (nitrogen and three carbons)
comprising the core of the NV center. The red dashed line denotes the
calculated total DOS of the lattice vibrations. An excellent quantitative agreement
between the experiments and the calculated projected DOS is clearly visible.
Both calculated DOS were smeared
by convolution with the Gaussian of the width 10~meV, 
corresponding to the low resolution of the experimentally determined DOS.}
\end{figure}

Our calculations allow a direct comparison with these experimental results.
First, let us focus of the DOS $g(\omega)$ calculated from the experimental
data in Ref.~\onlinecite{Davies}.
The qLVMs are expected to play primary role in optical transition, in comparison
with the delocalized phonons, and we expect that the experimentally extracted DOS
is associated mostly with qLVMs, as $g(\omega)$ is mostly governed by the 
vibration amplitude at the defect site \cite{Maradudin}.
In Fig.~\ref{fig8}, we compare the experimentally obtained
$g(\omega)$ curve (taken from Ref.~\onlinecite{Davies}) with the total vibrational DOS,
and with the projected DOS associated only with the vibrations of the nitrogen atom
and the three carbons adjacent to the vacancy. 
For comparison with the experimental
data, the discrete lines corresponding to individual modes have been broadened,
following the standard practice, by convolution with the Gaussian of the width 10~meV,
corresponding to the low resolution of the method used in Ref.~\onlinecite{Davies}.
The projected DOS practically coincides with the experimental data, thus confirming
our expectation, while the
total DOS demonstrates serious difference in the positions of the two peaks
and in their relative amplitude. Comparing Figs.~\ref{fig8} and \ref{fig3},
we see that the two experimentally found peaks correspond to the two groups
of qLVMs, one around 70~meV and the other around 140--160~meV. Note that the
high-frequency group of qLVMs has not been reported before.
These results clearly show the crucial
influence of the qLVMs on the optical emission and absorption.

In the subsequent work \cite{DaviesHamer}, the influence of vibrations on the
emission/absorption has been analysed in much finer detail, at the level
of individual vibronic peaks. It has been found that the vibronic spectra
show a clear one-phonon sideband with the vibronic frequency 
of approximately 70~meV. The experiments have been performed at low temperatures,
so that the starting state for both absorption and emission is likely to
contain no qLVM quanta, while the final state for absorption (emission)
will contain a single quantum of the qLVM corresponding to the excited (ground)
state \cite{Pekar,Nizovtsev01}. We see immediately that the qLVM group with the frequencies near 70~meV,
analyzed above and shown in Figs.~\ref{fig6} and \ref{fig7},
is exactly the right candidate for explanation of this feature. They also have
correct $a$ symmetry, see Tables~\ref{tab1} and \ref{tab2}.

A puzzling feature found in Ref~\onlinecite{DaviesHamer} is that the absorption
sideband looks like a doublet of two individual lines, with 
the width of order of 20~meV each, and the peak-to-peak distance of only 3~meV
(significant broadening is likely caused by the random stresses in the sample).
Note 
that the multiplet structure of the one-phonon sidebands is fully consistent
with our results: we have a group of at least three closely spaced qLVMs near
70~meV in both excited and ground state. The apparently single-peak sideband in emission and
doublet in absorption, shown in Fig.~6 of Ref.~\onlinecite{DaviesHamer},
are equally likely to be made of three peaks of different height, each inhomogeneously
broadened by 20 meV. Therefore, we do not analyze this
feature in more detail. Possibly, future measurements on a single NV center, free
of such a large inhomogeneous broadening, will convincingly resolve the
fine structure of these sidebands. 

For a similar reason, we do not discuss the spectral signatures of the
ground-state 41.66 meV and the excited-state 31.62 meV qLVMs. One might argue
that the steeply rising edge of the one-phonon sideband in Fig.~6 of Ref.~\onlinecite{DaviesHamer}
does show a shoulder in the right position (near 1.9 eV), but this claim
would be way too speculative.

\section{Conclusions}

We have performed detailed first-principles study of the localized distrortions
and (quasi)localized vibration modes (qLVMs) of NV centers in diamond. 
Our calculations show that the presence of NV center gives rise to a large
number of qLVMs with different degree of localization. 

We found that in the high-symmetry $C_{3v}$ excited state of the NV center
some qLVM become unstable,
and lead to lowering of the symmetry to $C_{1h}$. Such a lowering is expected,
since the excited state of the NV center is a typical example of Jahn-Teller
unstable system. Our results seem to support the earlier suggestion \cite{Fu09}
that the dynamical Jahn-Teller effect takes place. Our calculations indicate
the regime of the weak vibronic coupling, with the mode frequencies about an order
of magnitude larger than the Jahn-Teller energy. While the quantitative
first-principle theory of this effect for the NV center is lacking,
it constitutes an exciting topic for future work.

Next, we have analyzed and described some interesting qLVMs in the ground
and excited orbital states of NV center, and discussed the changes in the
qLVM density of states occuring upon optical excitation. We have performed
a comparison of our results with the experimental data reported in 
Refs.~\onlinecite{Davies,DaviesHamer}.
We have shown that the calculated DOS of the qLVMs is in very good quantitative
agreement with the results of Refs.~\onlinecite{Davies}, demonstrating
two peaks near 70 meV and 140--160 meV. These peaks are determined by the corresponding groups
of highly localized modes. Therefore, as expected,
qLVMs crucially impact the optical properties of the NV centers in 
diamond. 

We have also demonstrated
that the group of qLVMs with frequencies near 70 meV is likely to explain
the features of the one-phonon sideband observed in Ref.~\onlinecite{DaviesHamer},
possibly including their multiplet structure. However, detailed 
resolution of this issue requires further experimental and theoretical work,
such as high-resolution measurements of the absorption/emission spectrum 
on a single NV center, and highly precise first-principles calculations of the electron-phonon
coupling.

\acknowledgements
We thank R. Hanson, D. D. Awschalom, G. D. Fuchs, A. Gali, and N. Manson for
useful discussions on phonon properties of the NV centers.
This work was performed at the Ames Laboratory of US Department of Energy,
and was fully supported by the Department of Energy --- Basic Energy Sciences,
Division of Materials Science and Engineering, 
under Contract No. DE-AC02-07CH11358. 
Most calculations have been performed using the computers at the National Energy Research Supercomputing Centre (NERSC) in Berkeley.
J. Z. also acknowledges
support from China Scholarship Council (File No. 2009631039).

\appendix
\section{Atomic displacement patterns of several ground-state and excited-state qLVMs}

Below, we present the patterns of atomic displacements for several ground-state
and excited-state qLVMs having larges IPR.

\begin{figure}
\includegraphics[width=9cm]{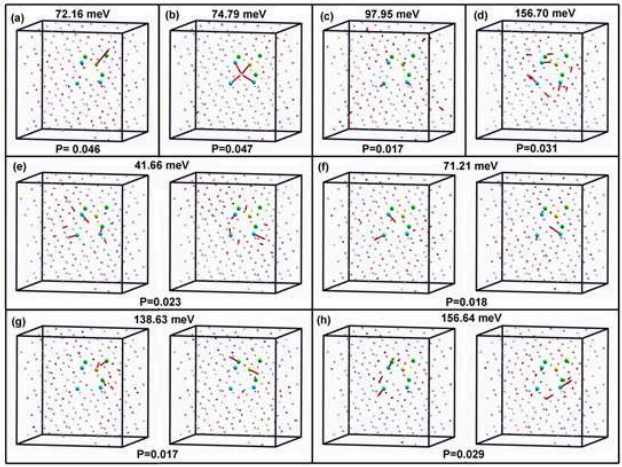}
\caption{\label{fig9} 
(Color online). qLVMs in the ground state of the NV center: atomic vibration patterns of  
several modes of different symmetry having largest IPR. Yellow spheres denote N atom, blue spheres denote the
C atoms adjacent to the vacancy, green spheres denote the C atoms adjacent to the N atom.
Small grey speres are the other carbon atoms in the diamond lattice.}
\end{figure}

In the $C_{3v}$ ground state, the lowest-energy localized $a$ mode at
72.16 meV is shown in  Fig.~\ref{fig9}(a). It is strongly
localized on the nitrogen: the N atom moves in the
direction perpendicular to the plane defined by the three C
atoms around the vacancy. The two $a$ modes shown in Fig.~\ref{fig9}(b)
and (c), with the frequencies 74.79~meV  and 97.95~meV are the
breathing modes of the four first-neighbor atoms of the
vacancy: all four atoms move to (or away from) the vacancy. 
The displacements of the four first neighbor atoms
in the 97.95~meV mode are smaller than those in 73.79~meV mode
indicating less localization of the breathing mode at
97.95~meV. The other $a$ mode at 156.7~meV, shown in Fig.~\ref{fig9}(d),
is a twisting mode around the NV center. The first- and second-neighbor carbon
atoms from the vacancy rotate in different directions. 

The eigenvectors of the four pairs of doubly degenerate $e$ modes, at
the frequencies of 41.66~meV, 71.21~meV, 138.63~meV, and
156.64~meV, are shown in Fig.~\ref{fig9}(e)--(h).
The two lower-energy localized $e$ modes, Fig.~\ref{fig9}(e) and (f), 
are mostly C-N-C angle bending modes. The third one, shown in Fig.~\ref{fig9}(g),
with a smaller IPR, is the N--C stretching mode 
at the NV center. The qLVM shown in Fig.~\ref{fig9}(h)
is a C--C stretching mode: carbon atoms around the
vacancy move in the out-of-plane direction, and at the same
time, their neighboring carbon atoms move in the opposite
direction. 

\begin{figure}
\includegraphics[width=9cm]{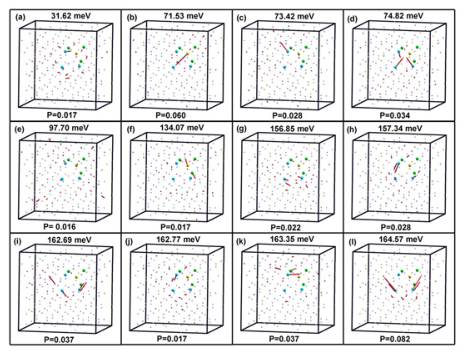}
\caption{\label{fig10} 
(Color online). qLVMs in the excited $C_{1h}$ state of the NV center: atomic vibration patterns of  
several modes having largest IPR. Yellow spheres denote N atom, blue spheres denote the
C atoms adjacent to the vacancy, green spheres denote the C atoms adjacent to the N atom.
Small grey speres are the other carbon atoms in the diamond lattice.}
\end{figure}

The displacement patterns of the qLVMs in the $C_{1h}$ 
excited state are shown in Fig.~\ref{fig10}, and are quite
different from the ground-state qLVMs. 
There is no doubly-degenerate mode in the structure with $C_{1h}$ 
symmetry. Among the twelve localized modes with largest IPR, 
only one, with frequency 71.53 meV, shown in Fig.~\ref{fig10}(b), is
similar to the ground-state mode at 72.16~meV. Six qLVMs
at 134.07~meV, 157.34~meV, 162.69~meV, 162.77~meV, 163.35~meV,
and 164.57~meV, shown in Fig.~\ref{fig10}(f,h,i,j,k,l) respectively,
are the N--C and C--C stretching modes. There
are three localized C--N--C angle bending modes at 31.62~meV,
Fig.~\ref{fig10}(a), at 74.82~meV, Fig.~\ref{fig10}(d), and 
at 156.85~meV Fig.~\ref{fig10}(g). One breathing mode is 
seen at 97.70~meV, Fig.~\ref{fig10}(e), which has smaller IPR.
Finally, the mode at 73.42~meV has a C atom moving opposite 
to the vacancy, as shown in Fig.~\ref{fig10}(c). In
all the excited-state qLVMs discussed
above, the C$_1$ carbon atom moves independently, while the
other two equivalent carbons move collectively to maintain the
$C_{1h}$ symmetry. The carbon atoms adjacent to the displaced
carbons move in the same plane. This plane is formed by the three
first-neighbor carbons in such a way that the rotation and translation
of the whole system is absent. Due to
the low $C_{1h}$ symmetry, all these modes are IR active.


\begin{thebibliography}{99}
\bibitem{SymmetryNote} For ease of comparison with the previous
works, we always keep the notations of the parent group, $C_{3v}$, when describing
the excited state. E.g., we denote the orbital excited state as having $^3E$ symmetry, 
although the $C_{1h}$ point group has only singlet representations.
\end{thebibliography}
\end{document}